# Structural and magnetic properties of $CoTeMoO_6$ revisited

Yu Li [1,2], Jared Coles [1,3], Xin Gui [4], Hyowon Park [1,5], Yan Wu [6], Xinglong Chen [1], Jing-han Chen [2], Xiaoping Wang [6], Huibo Cao [6], Shane Stadler [2], Omar Chmaissem [1,3], David P. Young [2], Stephan Rosenkranz [1], John F. DiTusa [2,7]

[1] Materials Science Division, Argonne National Laboratory, Lemont, Illinois 60439, USA

[2] Department of Physics & Astronomy, Louisiana State University, Baton Rouge, Louisiana 70803, USA

[3] Department of Physics, Northern Illinois University, DeKalb, IL 60115, USA

[4] Department of Chemistry, University of Pittsburgh, Pittsburgh, PA 15260, USA

[5] Department of Physics, University of Illinois at Chicago, University of Illinois, Chicago, Illinois 60607, USA

[6] Neutron Scattering Division, Oak Ridge National Laboratory, Oak Ridge, Tennessee 37831, USA

[7] Department of Physics, School of Science, Indiana University Indianapolis, Indianapolis, Indiana 46202, USA

We have conducted a comprehensive investigation into the magnetic properties of the chiral multiferroic material $CoTeMoO_6$. In contrast with the previous claim of canted antiferromagnetic order with ferromagnetic components [1], our investigation reveals an antiferromagnetic ground state with compensated moments, providing an interesting platform for exploring exotic material properties. Through careful measurements of magnetization under a series of applied field, we demonstrate that there exist two sequential field-induced magnetic transitions in $CoTeMoO_6$, with one occurring at $H_{c1} = 460$ Oe along the a-axis, and the other at $H_{c2} = 1.16$ T with the field along the b-axis. The values of $H_{c1}$ and $H_{c2}$ exhibit strong angular dependence and diverge with different rates as the applied field is rotated 90 degrees within the ab plane. This reflects the distinct nature of these transitions, which is further supported by the different critical behavior of $H_{c1}$ and $H_{c2}$, characterized by the values of $\gamma$, in the function of $H_c = H_0 \times (1 - T/T_c)^{\gamma}$. Furthermore, we have demonstrated that there exist structural and magnetic twin domains in $CoTeMoO_6$ that strongly affect the experimental measurement of their macroscopic properties. Intriguingly, these twin domains can be related to the orthorhombicity/chirality of the crystal structure with the space group $P2_12_12$. We further explored the magnetic and structural domains with uniaxial pressure and polarized light microscopy. Our results suggest that $CoTeMoO_6$ could be used as a unique platform for investigating the intriguing physics involving intertwined degrees of freedom. The tunability of the underlying domain distribution and its strong anisotropy could also be useful for developing functional devices and applications.

## 1. Introduction

Driven by significant advancements in fundamental physics and promising applications in information technology, substantial progress has been made in the past decades in discovering and characterizing exotic magnetic states, and many innovative concepts beyond the existing understanding have been proposed [2,3,4,5,6,7,8,9,10]. As a powerful tool, the analysis of symmetry classification [8,9,10,11,12,13] has been extensively used in predicting and studying potential materials of interest. For example, traditional wisdom considered ferromagnetism and antiferromagnetism as the two basic magnetic phases. It is now realized that a third magnetic phase could exist, called *alter-magnetism* [14, 15]. Alter-magnets exhibit both time-reversal symmetry-breaking, characteristic of ferromagnets, and antiparallel magnetic order with vanishing net magnetization, a feature of antiferromagnets. Since it was proposed, the concept of alter-magnetism has attracted significant interest from both theorists and experimentalists [16, 17, 18]. Another example is the discovery of magnetic skyrmions in MnSi [19]. At zero field, the magnetic ground state of MnSi is a helical order with a very long pitch $\lambda \sim 180$ Å [20]. In the presence of a finite magnetic field, magnetic skyrmions can arise as swirling topological defects and host exotic transport behaviors [21, 22]. It is found that the emergence of magnetic skyrmions in MnSi is intimately linked to the existence of the Dzyaloshinski-Moriya (DM) interaction [23,24] due to its non-centrosymmetric chiral structure. It has been established that such magnetic materials with chiral space groups could be potential hosts of exotic spin textures. Among these materials, $CoTeMoO_6$ [1, 25, 26, 27, 28] stands as a promising candidate that is yet to be fully explored.

$CoTeMoO_6$ is an important member of the telluromolybdate family $ATeMoO_6$ and was first synthesized in 1975 [25]. This family has been extensively studied for its promising catalytic [29, 30] and nonlinear optical [31, 32] properties. The crystal structure of $CoTeMoO_6$ consists of two-dimensional layers of distorted $CoO_6$ octahedra separated by $MoO_4$ tetrahedra and $TeO_4$ polyhedra. The presence of a $d^0$ transition metal ($Mo^{6+}$) and a lone-pair cation ($Te^{4+}$) has contributed to the second-order Jahn-Teller effect and are thought to be responsible for its strong nonlinear optical (NLO) properties [33]. Recently, there has been revived interest in $CoTeMoO_6$ due to its intriguing magnetic behaviors and potential functionalities [28]. Specifically, these interests arise from two special features hosted by its crystal symmetry. First, the space group of $CoTeMoO_6$ is chiral, so the DM interaction exists and might result in the formation of possible nontrivial spin textures [19, 34, 35, 36, 37, 38]. Second, the structural unit cell of $CoTeMoO_6$ contains two magnetic ions. In the presence of suitable symmetry and antiparallel magnetic order between the two sub-lattices, alter-magnetism could be realized [16]. In order to address these problems, a comprehensive investigation on the magnetic properties of $CoTeMoO_6$ is highly desired.

A previous study [1] on $CoTeMoO_6$ suggested that the magnetic ground state is antiferromagnetic with canted ferromagnetic components. According to this model, the major components of the ordered moment are antiparallel and aligned along the b-axis. Meanwhile, there are small components titling towards the a-axis, resulting in a ferromagnetic response indicated by low-temperature magnetization measurements. However, the existence of finite ferromagnetism at low temperatures was not confirmed by powder neutron diffraction experiments and is also inconsistent with the magnetic structure of the isostructural compound $MnTeMoO_6$ [39]. To further understand the situation in $CoTeMoO_6$, we performed a careful examination of the magnetic properties with the Magnetic Property Measurement System (MPMS), neutron/X-ray diffraction, and polarized light microscopy. We demonstrated that the actual magnetic ground state in $CoTeMoO_6$ is an antiferromagnetic order with zero net magnetization. Furthermore, we show that there exists a field-induced magnetic phase separated from the zero-field antiferromagnetic state with a very small critical field, ~ 460 Oe, when the magnetic field is applied along the a-axis. This is reminiscent of the case in MnSi, where there is a small critical field separating the zero-field helical phase from the partially polarized conical magnetic state. Because magnetic skyrmions can arise in MnSi by tuning the system across the phase boundary [40,41] with the magnetic field, it would be of particular interest to investigate the evolution of possible spin texture in $CoTeMoO_6$ by preparing the sample with different magnetic fields. In addition, we also found strong in-plane magnetic anisotropy in $CoTeMoO_6$, which has been overlooked in previous investigations, probably due to the existence of magnetic/structural twin domains. Utilizing uniaxial pressure and polarized light microscopy, we demonstrated that the magnetic, structural, and optical properties are intimately coupled. More intriguingly, these properties could be related to the orthorhombicity/chirality of the crystal structure. This provides a unique avenue for investigating the inter-coupling among multiple degrees of freedom and exploring useful functionalities with external uniaxial pressure.

## 2. Sample preparation and experimental methods

Polycrystalline samples of $CoTeMoO_6$ were synthesized using a conventional solid-state reaction method. High-purity powders of CoO, $TeO_2$ and $MoO_3$ were weighed to be in a stoichiometric ratio and mixed thoroughly. The resulting mixture was ground and pressed into pellets, which were then sealed in an evacuated quartz tube and kept at 560 ℃ for 24 hours. After the reaction, the pellets were ground and sintered for a second time, resulting in a single-phase $CoTeMoO_6$ powder with a purple-blue color, as confirmed by powder X-ray diffraction (XRD).

To obtain single crystals of $CoTeMoO_6$, polycrystalline samples were mixed with additional flux [26] of $TeO_2$-$MoO_3$ mixture (1:0.8) and heated in a quartz tube to 700 ℃. The concentration of $CoTeMoO_6$ in the mixture is about 16% in mole ratio. The mixture was slowly cooled down to 500 ℃ over several days. While reactions with the quartz tubes were observed on the surface, large plate-like single crystals with a typical size of ~ 5mm in diameter could be easily cleaved from the final product.

Room-temperature powder X-ray diffraction (XRD) was performed using a Scintag XDS2000 powder diffractometer with Cu Kα radiation. Single crystal XRD was conducted on a Bruker D8 VENTURE equipped with Mo $K_\alpha$ radiation. Several crystals were picked and tested separately to ensure reproducibility. Indexing, space group selection and numerical absorption correction were performed using XPREP implemented in APEX4. The crystal structure was solved using the SHELXTL package. The solved crystal structure is shown in Fig. 1(a) and (b), and the refinement of single crystal XRD is shown in Table. 1-3.

Temperature- and field-dependent *dc* magnetization and *ac* susceptibility measurements were conducted using a Quantum Design (QD) Magnetic Property Measurement System (MPMS). To measure the magnetization of the single crystals in different orientations, we designed and made a rotation device using G-10 glass-epoxy laminates that were small enough to be compatible with the MPMS measurement system. This device allowed for sample orientation control within an accuracy of 2 degrees. For uniaxial pressure measurements, we manufactured a detwinning device with G-10 epoxy. We note here that, in order to change the orientation of the sample or modify the applied pressure, the sample has to be warmed up to room temperature and removed from the instrument.

In the uniaxial-pressure measurements, the single crystals were cut into a rectangular shape with the edges along the crystalline a- and b-axis. Meanwhile a naturally cleaved edge is often seen along the diagonal direction.

Polarized light microscopy measurements were performed by placing a linear polarizer between the light source and the sample under an optical microscope. While the direction of the light polarization was undetermined in the configuration, it is fixed during the sample rotation. Therefore, the relative angle between the light polarization and the crystalline axes varies in a series of observation.

Time-of-flight single crystal neutron diffraction was performed on TOPAZ using the wavelength-resolved Laue technique at the Spallation Neutron Source (SNS) at Oak Ridge National Laboratory (ORNL). Data were collected at 295 K and 5 K with the conditions listed in Table 4. The data reduction and integration of Bragg intensities were conducted on the MANTID platform [42]. The nuclear and magnetic structures were refined using JANA2020 [43].

We have adopted the density functional theory plus U (DFT+U) method to perform the magnetic band structure calculations of $CoTeMoO_6$. For DFT+U calculations, we used the Vienna Ab-initio Simulation Package (VASP) code [44, 45] along the Perdew-Burke-Ernzerhof (PBE) functional for the exchange-correlation energy of DFT. We also used a $12 \times 12 \times 8$ k-mesh with the energy cutoff of 400 eV for the plane-wave basis to perform the DFT calculation based on the experimental crystal structure of $CoTeMoO_6$. We used the Hubbard interaction U= 2.5 eV and 5 eV, and the Hund's coupling J = 0.8 eV to treat the strong correlation effect of the Co $d$ orbitals.

## 3. Experimental Results and Data Analysis

### A. Magnetization and susceptibility

Figure 1 (d) displays the measured magnetization of $CoTeMoO_6$ as a function of field at T = 4 K with the field applied along the a- and b-axis. Both data displays small hysteresis loops at low field (close to zero field) and a step-like feature around H ≈ 1.2 T, indicating a field-induced magnetic transition. However, there is a discrepancy in the magnitude between the two directions, indicating the existence of in-plane magnetic anisotropy. Our further measurements and analysis reveal that this difference actually reflects the existence of magnetic twin domains which are not equally populated. In order to study the intrinsic behavior, we extract the magnetization for a single domain by assuming that the measured curves are contributed by twin domains whose main crystalline a-axis (or b-axis) are perpendicular to each other. Accordingly, the relationship between the intrinsic magnetization, $M_a$, $M_b$, and $M_c$, for a single domain, and the magnetization, $M_a^*$, $M_b^*$, and $M_c^*$, measured from a sample with twin domains, can be described as below:

$$\begin{cases} M_a(H,T) \times (1-p) + M_b(H,T) \times p = M_a^*(H,T) \\ M_a(H,T) \times p + M_b(H,T) \times (1-p) = M_b^*(H,T) \\ M_c(H,T) = M_c^*(H,T) \end{cases} \quad \text{(Eq. 1)},$$

where $M_x$ denotes the magnetization measured with the field applied along the x-direction, $H$ and $T$ are the measured magnetic field and temperature, and $p$ and $1-p$ are the volume fractions of minority and majority domains. We found that the value of $p$ changes significantly for different pieces of crystals, from ~ 0.16 to about half. In Fig. 1 (e) and (f), we plot the extracted magnetic hysteresis and zero-field-cooled (ZFC) magnetization curves for single-domain along the a-, b- and c-axis at 4

K. We find that the intrinsic magnetization for a field along the a-axis, $M_a$, still shows the magnetization step near zero field, though slightly enhanced as compared to the $M_a^*$ in Fig.1 (d), but the step observed in $M_a^*$ at a higher field ≈ 1.2 T is entirely absent in the extracted $M_a$ from Eq. 1, as shown in Fig. 1(f). Conversely, the extracted intrinsic $M_b$ does not show the step near zero field but only the step at $H \approx$ 1.2 T. This suggests that the magnetization step around zero field is actually the intrinsic magnetic response of a single domain for a field along the a-axis, while the step at H ≈ 1.2 T is the response to a field applied along the b-axis. This is consistent with both our assumption about the twinning effect, and the existence of strong in-plane magnetic anisotropy. In comparison to the measurements in the ab plane, the magnetization is dramatically reduced in magnitude when the field is along the c-axis. This suggests that the c-axis is essentially the hard axis of this magnetic system.

To further explore the magnetism in $CoTeMoO_6$, we measured the temperature-dependent susceptibility and further calculated the intrinsic single-domain response along the three main crystalline axes. The temperature-dependent magnetic susceptibility, $M_a/H$,$M_b/H$, and $M_c/H$, are shown in Fig. 2 (a) and their inverses are plotted in Fig. 2 (b). A magnetic transition is clearly observed at $T_N = 25\ K$. Cure-Weiss fits of the magnetic susceptibilities above 100 K yield different values for the three crystalline directions, with $\mu_{eff,a} = 5.0\,\mu_B/Co^{2+}$, $\theta_{CW,a} = -54\ K$ for the a-axis, $\mu_{eff,b} = 5.4\,\mu_B/Co^{2+}$, $\theta_{CW,b} = -47\ K$ for the b-axis, and $\mu_{eff,c} = 3.9\,\mu_B/Co^{2+}$, $\theta_{CW,c} = -109\ K$ for the c-axis, respectively. The average of these estimated moments is consistent with the previous result from polycrystalline samples [1]. It is worth noting that the average magnetic moment is significantly larger than the expected value of ∼3.88 $\mu_B/Co^{2+}$ for high-spin state of $Co^{2+}$, indicating the existence of unquenched orbital angular momentum.

In Fig. 3 (a), we present a series of magnetization measurements along the nominal a-axis after a zero-field-cooling procedure and focus on the low field region from zero to 0.6 T. While these data have not been corrected for a single domain, they clearly show that in the low-field region below 500 Oe, there exists an antiferromagnetic state with vanishing net magnetization. This is in sharp contrast to the previous claim that finite ferromagnetism was observed at low temperatures. While nearly zero magnetization can also be observed in some soft ferromagnets with balanced magnetic domains, we note that the step at ~ 500 Oe is sharp and persists up to 23 K, very close to the magnetic transition temperature. We therefore conclude that rather than a redistribution of existing magnetic domains, the feature around 500 Oe is actually related to a field-induced magnetic transition which separates a zero-field antiferromagnetic state from a partially polarized phase with finite magnetization.

This argument is further supported by the measurement of the magnetic hysteresis at a series of temperatures as shown in Fig. 3 (b). While our data are similar to those previously reported [1, 28], we find that the magnetization at 4 K starts to drop around 500 Oe as the field decreases and develops an anomaly as indicated by the arrows at the same field magnitude on the opposite side of the y-axis at H = -500 Oe. The same behavior can also be found in the lower branch of the hysteresis as the field ramps up. Furthermore, we find that the hysteresis loop transforms into a wasp-waist shape at elevated temperatures and finally diminishes as sharp steps at ±500 Oe on both sides of the y-axis. This behavior can be attributed to an intrinsic interaction in this material with a magnitude equivalent to the characteristic field of H ≈ 500 Oe. This interaction is essentially responsible for the formation of antiferromagnetic ground state and keeps the system away from other magnetic phases with finite magnetization. This is very similar to the effect of the DM interaction which [20, 46] transform pristine ferromagnetism into a helical order in MnSi.

Moreover, the magnetic hysteresis at 4 K implies that the system can be trapped in a metastable magnetic state with finite magnetization under field cooling. Nevertheless, the energy barrier separating this metastable state from the ground state can be overcome by thermal fluctuations at higher temperatures where the system goes back to its antiferromagnetic state with vanishing magnetization at zero field. This is in agreement with the evolution of the magnetic hysteresis as the temperature increases as shown in Fig. 3 (b). It is still unclear how the magnetic domains change with variations in field and temperature, and whether the DM interaction in this material could lead to the emergence of nontrivial spin textures [37,38].

To understand the possible magnetic anisotropy in this material, we conducted a systematic magnetization measurement at a series of field directions, as shown in Fig. 4 (a). The measurements were performed at 2 K after ZFC for each field direction. Since our primary interest lies in the values of critical fields at different angles, the presented data have not been corrected for single domain analysis and are presented in arbitrary unit. For most angles, we observe two clear steps in the magnetization curves – one at a small field denoted as $H_{c1}$ and the other at a higher field labeled as $H_{c2}$. By calculating the first derivative, $dM/dH$, we further estimated the values of $H_{c1}$ and $H_{c2}$, as shown partly in Fig. 4(b) for selected angles in the low field region. Alongside the primary peaks, we observe secondary peaks with reduced heights, suggesting additional contribution. In Fig. 4 (c), we present magnetization curves in a series of directions nearly perpendicular to those angles in Fig. 4 (b). A second magnetization step is clearly visible with distinct angle dependence and is further supported by the calculated $dM/dH$ curves in Fig. 4 (d), indicating possible existence of two domains. In Fig. 4 (e), we summarized the angle dependence of $H_{c1}$ for both the primary and secondary peaks in the $dM/dH$ curves. We found that while their minima are located at nearly perpendicular directions, their angle dependence exhibit similar behavior. This can be perfectly described by a fitting function of $H_{c1} = H_{c1,0} \times \cos^{-1}(\theta)$, where $\theta$ represents the angle between the field direction and the crystalline a-axis. The value of $H_{c1,0}$ is estimated to be 460 Oe. The success of this fitting procedure strongly supports the existence of two perpendicular domains and that the critical field $H_{c1}$ for a single domain has a minimum along the a-axis. This kind of angle dependence is analogous to that observed in $HoNi_2B_2C$, which arises from the difference in free energy of the two relevant magnetic states [47]. Consequently, we attribute the primary peaks in the $dM/dH$ curves to the transition of majority domains at $H_{c1}$ and the secondary peaks to minority domains, labeled as domain 1 and domain 2 respectively in Fig.4 (e).

In Figure 4 (f), we also summarize the angular dependence of $H_{c2}$ using the same approach as for $H_{c1}$. Unlike $H_{c1}$, the minimum of $H_{c2}$ for the corresponding domain lies along the b-axis. Additionally, we find that $H_{c2}$ varies at a rate significantly faster than $H_{c1}$, as the field approaches the diverging direction. Our analysis further indicates that the function $H_{c2} = H_{c2,0} \times \cos^{-4.5}(\theta)$ provides a good fit for the data, with the exponent of -4.5 being considerably larger than that for $H_{c1}$. These discrepancies in the angle dependence suggest that magnetic transitions at $H_{c1}$ and $H_{c2}$ originate from distinct mechanism.

To account for the distinct nature of $H_{c1}$ and $H_{c2}$, we present the extracted magnetic phase diagram of $CoTeMoO_6$ as a function of field (H) and temperature (T) along the a- and b-axes, in Fig. 5 (a) and (b), respectively. Presumably, in an ideal crystal with a single domain and perfect alignment, only one of the two transitions, $H_{c1}$ and $H_{c2}$, would be observable along either a- or b-axis, as the other would diverge and be unobservable. In practice, due to the finite mosaic of crystals and imperfect field orientation, the divergent behavior is replaced by a smearing of the transition. Furthermore, to fit the data in Fig. 5 (a) and (b), we use the function $H_c = H_0 \times (1 - T/T_c)^{\gamma}$ , in which $T_c = 25$K is the transition temperature at zero field and $H_0$ is the critical field at absolute zero temperature. While accurate determination of the critical exponent is only feasible in the vicinity of $T_c$, the estimated values of $\gamma$ differ significantly for $H_{c1}$ and $H_{c2}$, with a value of 0.13 and 0.36, respectively. This striking disparity strongly supports that the magnetic transitions at $H_{c1}$ and $H_{c2}$ are driven by different mechanisms.

### B. Single Crystal Neutron diffraction

In order to understand the magnetic structure in $CoTeMoO_6$ under zero field, we conducted neutron diffraction measurements at both 295 K and 5 K and present the refined results in Fig.6 (a). In agreement with our temperature-dependent magnetic susceptibility measurements, we observed antiferromagnetic reflections at 5 K in addition to the Bragg peaks of the orthorhombic crystal structure at 295 K. These magnetic reflections can be indexed with a magnetic propagation vector $k_{22} = (0,0,1/2)$. In Fig. 6 (b), we show the solved magnetic structure from our refinement at 5 K. Our analysis reveals that the magnetic moments in $CoTeMoO_6$ are mostly aligned along the b-axis, with small components titling towards the a-axis. The tilting angle is nearly double of that observed in previous powder neutron diffraction measurements. Although the in-plane antiferromagnetic wave vector is $q_{ab} = (0,0)$, there is a finite c-axis component with $q_c$ = ½, requiring the doubling of the c-axis and an antiferromagnetic arrangement between neighboring layers.

In Fig. 7 (a) and (b), the doubling of the magnetic unit cell is evident from the presence of magnetic peaks at $(h,k,l)$ with $h,k = n$, and $l = n + 0.5$, where n is an integer. No extra magnetic peaks were seen in the [H, K] plane with L equal to an integer, precluding the existence of ferromagnetic correlations along the c-axis. Group theory representation analysis performed with Jana2020 identified four possible magnetic space groups: $P_c 2_1 2_1 2$, with the origin shifted to (0,0,0) ($mZ_1$) or to (0,0,1/4) ($mZ_2$), and $P_c 2_1 2_1 2_1$, with the origin shifted to (1/4,0,0) ($mZ_3$) or to (1/4,0,1/4) ($mZ_4$). The first two magnetic space groups ($mZ_1$ and $mZ_2$) allow magnetic moments for the Co along the c-axis, while the last two space groups allow magnetic moments in the $ab$ plane with non-zero $M_x$ and $M_y$ components. Refinements using all four models showed that the best fit was achieved with the $mZ_4$ space group, as detailed in Table 5. In Table 6 we show a list of selected bond-lengths and bond-angles at 295 K and 5 K.

In Fig. 7 (c) and (d), we investigate the temperature dependence of nuclear and magnetic peak intensities at several representative locations with Q = (0,0,3), (0,1,4.5) and (0,0,3.5). In Fig. 7 (c), a steep step around 60 K is clearly seen, in accordance with previous X-ray diffraction measurements [28]. Raman spectroscopy suggests that this anomaly reflects the change of local structure and might be related to the displacement of oxygen. At temperatures around 20 K, another anomaly is observed in Fig. 7 (c), likely associated with the establishment of a long-range magnetic order seen in Fig. 7 (d). These observations indicate strong coupling between the crystal structure and magnetism in this material. In addition, the intensity at Q = (0,0,3.5)/ (0,1,4.5) in Fig. 7 (d) starts to increase as the temperature decreases in the range between 20 K and 60 K. This suggests that short-range magnetic correlations exist far above the magnetic transition temperature. It would be interesting to study these short-range correlations in single crystals with diffuse neutron scattering.

In addition, we also observed extra peaks, with negligible intensities, in the vicinity of several main structural Bragg peaks at 295 K. We attribute these peaks to the Bragg peaks from minority domains. The estimated volume fraction, $p$, of the minority domains from the neutron diffraction measurement is significantly smaller than the values obtained from the magnetization measurement. However, this is not entirely surprising given that the population of domains relies on the history of experienced strain. This is further revealed by the uniaxial pressure measurements which are in the following section. The observation of twin domains in single crystal neutron diffraction is consistent with our magnetization measurements at low temperatures. However, these peaks cannot significantly affect our magnetic structure refinement due to the weak intensities and the advantage of single crystal diffraction.

### C. Twin domains and uniaxial pressure effect

The presence of unbalanced magnetic/structural twin domains observed in our magnetic measurements and neutron diffraction experiments is intriguing and warrants further exploration. In systems with magnetoelastic coupling, it is generally thought that the magnetic properties can be influenced by external pressure or strain. For instance, in iron pnictide superconductors [48, 49, 50], the magnetic moments are linked to the structure of the so-called nematic phase. By applying a uniaxial pressure along one of the crystalline axes, it becomes possible to suppress one domain and favor the other, resulting in a detwinning effect. Inspired by this idea, we designed and fabricated a uniaxial pressure device [Fig. 8 (a)] to investigate the detwinning effect in $CoTeMoO_6$, as shown in Fig.8 (b). We note here that we are not aware of the values of the applied pressure and have no intention of providing an estimate with the current device.

In Fig. 8 (c), we plot the magnetization curve with and without applied uniaxial pressure after subtracting the magnetic background from an empty device. Since the magnetic transitions at $H_{c1}$ and $H_{c2}$ in a single measurement can be attributed to different domains, it is convenient to use the magnetization steps at each transition, $\Delta M_{c1}$ and $\Delta M_{c2}$, as indicators of the domain population. From the measurement without applied pressure (blue curve), it is clear that $\Delta M_{c2}$ is significantly larger than $\Delta M_{c1}$, indicating an unbalanced population of domains, with the majority having their b-axis parallel to the field direction. In the second measurement (red curve), because the pressure is applied along the shorter b-axis of the majority domain, there is only slight change in $\Delta M_{c1}$ and $\Delta M_{c2}$, mainly due to the reduction of the minority domain, as shown in Fig. 8 (c).

In Fig. 8 (d), we selected a piece of crystal with nearly equal populations of the two domains, indicated by the comparable magnitudes of $\Delta M_{c1}$ and $\Delta M_{c2}$. After applying uniaxial pressure, the magnetization curve undergoes a dramatic change, demonstrating the detwinning effect. However, it should be mentioned that this sample is not fully detwinned, as a finite $\Delta M_{c1}$ is observed in Fig. 8 (d), while it is supposed to be zero in a fully detwinned sample. We then repeated the measurement after relieving the pressure and found that the magnetization curve does not return to its initial state. This finding aligns with the previous observation of varied domain populations in different pieces of crystals, indicating that the detwinning effect is irreversible at room temperature, exhibiting history-dependence. In order to investigate whether there is a tetragonal-to-orthorhombic transition at temperatures above room temperature, we performed differential scanning calorimetry (DSC) measurements but did not find evidence of such a structural transition in $CoTeMoO_6$ up to its melting/decomposition temperature. Our investigation using a uniaxial pressure device clearly demonstrates the close relationship between magnetic and structural domains whose distributions can be controlled by applying external strain.

To directly observe the structural domains in real space at room temperature, we performed an optical microscopy measurement with a light polarizer placed between the light source and the sample. Fig. 9 shows a few representative optical images, in which stripe-like patterns with alternating colors are observed, reminiscent of stripe domains found in iron pnictides [51]. These stripes are formed at an angle of approximately 45 degrees away from either the a- or b-axis of the orthorhombic crystal structure. Although the light polarization was undetermined in our setup, we discovered that the optical response of these stripe domains exhibited a 2-fold symmetry during sample rotation. For example, in Fig. 9 (a), the majority domains display a light purple color and are nearly transparent, while the minority domains exhibit a pink color. As the sample is rotated by 45 degrees, the contrast between these domains gradually diminishes and completely disappears in Fig. 9 (b). With further rotation, at $\omega = 90^{\circ}$ in Fig. 9 (c), the contrast between the two domains is reversed, with the majority area exhibiting a pink color and the minority area becoming transparent. This behavior is typical of birefringent or refractive materials, which have extensive applications in optical devices and industrial technology [52, 53, 54]. Our observation of birefringence agrees with a previous study on polarized absorption spectra [26], which reported an absorption peak in the visible light wavelength region (585-583 nm) and observed different intensities when the electric field was changed from the a-axis to the b-axis. While our observations provide direct evidence for the existence of orthorhombic domains at room temperature and indicate their birefringent properties, they also suggest an intriguing coupling among structural, magnetic, and electronic degrees of freedom, calling for further experimental investigations using uniaxial pressure.

## 4. Discussion and Conclusion

### a. Alter-magnetism

The unique properties of alter-magnetism, such as momentum dependent spin splitting in reciprocal space and zero net magnetization within the unit cell in real space, have generated considerable interest [14, 15, 16, 17, 18]. To achieve this magnetic state, a structural unit cell must contain an even number of magnetic ions with opposite spins and exhibit a zero-q antiferromagnetic order. $CoTeMoO_6$ satisfies these requirement as it contains two magnetic ions, $Co^{2+}$, in its structural unit cell. A recent symmetry analysis [16] predicted that $CoTeMoO_6$ could host alter-magnetism with an antiferromagnetic structure within the ab plane and ferromagnetic alignment along the c-axis. While our single crystal neutron diffraction has excluded this possibility, it remains intriguing to explore whether other magnetic materials with similar crystal structures may achieve alter-magnetism.

While experimental investigation [55,56,57] of the conjectured magnetic structure within $CoTeMoO_6$ is unattainable, spin-polarized DFT calculations can offer valuable insight into the band structure of such a hypothetical magnetic state. Previous research suggests that DFT calculations on alter-magnetism are not greatly affected by the inclusion of relativistic spin-orbit coupling and correlation effects [14]. Therefore, performing spin-resolved DFT calculations on the band structure of $CoTeMoO_6$ using the conjectured magnetic structure may help identify potential candidates for alter-magnetism.

We have conducted DFT calculations based on the crystal structure determined by X-ray measurements. The total energy for an antiferromagnetic state is found to be 34 meV lower than that of the ferromagnetic order, consistent with the compensated magnetic state. To account for the interaction of Co $d$ orbitals, we also performed DFT+U calculations with U values of 2.5 eV and 5 eV. In Fig. 10, we plot the band structure from our DFT+U calculation with U = 5 eV, showing significant spin splitting along the G-M direction in the 2D Brillouin zone. Remarkably, the observed band splitting remains consistent across all our calculations, regardless of U values. In contrast with the noncollinear magnetic structure calculated in Ref [28], the band splitting in the collinear antiferromagnetic state suggests that possible alter-magnetic state could be achieved by tuning the magnetic exchange couplings and thus the magnetic ground states. Therefore, it could be a promising avenue to explore magnetic materials with similar structures by tuning the relevant magnetic interactions through chemical substitution.

### b. Complexity of multiple interactions and degrees of freedom

The magnetism of $CoTeMoO_6$ becomes even more intricate when twin domains are taken into consideration. On one hand, the orthorhombic structure inherently gives rise to a set of twin domains with four different orientations associated with different shear distortion of a primitive square lattice, as illustrated in Fig. 11 (a). On the other hand, due to the chiral space group of $P2_12_12$, which has two $2_1$ screw axes, there are two enantiomers corresponding to each orthorhombic unit cell, as illustrated in Fig. 11 (b). In principle, both type of domain structures could impact the low-temperature magnetic behavior through single-ion anisotropy or the DM interaction, further complicating the twinning problem. As a consequence, it is still unclear which factor dominates and controls the low-energy magnetic behavior of $CoTeMoO_6$.

Our single crystal neutron diffraction experiments have confirmed that the nearest neighboring antiferromagnetic interaction is the strongest interaction, dictating an antiferromagnetic order in the system. In combination with magnetic anisotropy and the DM interaction, an antiferromagnetic order emerges with moments primarily along the b-axis and small canted components pointing towards the a-axis. While these canted in-plane components may contribute to a polarized magnetic state, they are effectively cancelled out by the antiparallel arrangement between neighboring layers. This is primarily due to the small inter-layer coupling, $J_c$, which is responsible for the c-axis doubling of the magnetic unit cell. As we mentioned in the previous section, it might be possible and interesting to tune the system into an alter-magnetic state by changing the sign of $J_c$ via chemical substitution.

Furthermore, the underlying mechanism of the field-induced transition is intriguing as it involves multiple relevant energy scales, such as magnetic anisotropy, the spin-orbit coupling and inter-layer antiferromagnetic interaction, $J_c$. For example, assuming that the magnetic transition at $H_{c1}$ is a spin-flip transition of the canted moments [indicated by the arrow in Fig. 1 (c)] in the presence of strong single-ion anisotropy, there could be two possible mechanisms. On one hand, when the applied field overcomes $J_c$, a field-polarized state can be attained by flipping half of the $Co^{2+}$ moments, leading to a ferromagnetic arrangement between neighboring layers. Alternatively, when the magnetic moment is strongly coupled with the local crystal structure in the presence of spin-orbit coupling, the transition can also be achieved by reversing the sign of DM interaction, as illustrated in Fig. 11 (b), and thus changing the tilting angle of canted moments, while inter-layer correlations remain mostly antiferromagnetic. In order to determine which mechanism is responsible for the transition at $H_{c1}$, single crystal neutron diffraction measurements under a magnetic field are essential and necessary for gaining further insights.

In addition, the significant magnetic hysteresis observed at low temperatures is also intriguing. This indicates that a large volume fraction of the material can be trapped in a metastable state with finite magnetization at the base temperature. The trapping may be associated with the pinning effect of local defects or global topological protection. While the dominant factor remains uncertain, further investigation to explore the history-dependent spin texture in $CoTeMoO_6$ would be very interesting. It is possible that the DM interaction plays a role in the evolution of the underlying spin texture, resulting in exotic magnetic behaviors.

And finally, despite strong experimental evidence connecting the low-temperature antiferromagnetic order to the room-temperature crystal structure, the existence of a structural change around 60 K makes this relationship ambiguous. A previous Raman spectroscopy study has identified the existence of stretch/contraction of oxygen-cation bonds within this temperature range. Our analysis of the crystal structure suggests that the distortion of certain bonds, specifically the bonds between Co and O, is responsible for the transformation between enantiomers with opposite chirality. However, experimental evidence for the existence of enantiomers and their transformation is currently lacking, and their relationship with low-temperature magnetism requires further experimental investigations using advanced X-ray diffraction techniques.

In summary, we conducted a comprehensive investigation of the structural and magnetic properties of $CoTeMoO_6$ by tuning temperature, field, uniaxial pressure, and light polarization. In contrast to the previously assumed canted antiferromagnetic order with ferromagnetic components, our findings demonstrate that the magnetic ground state is, in fact, an antiferromagnetic order with compensated magnetization. These results encouraged us to explore possible exotic magnetic phases in $CoTeMoO_6$, such as alter-magnetism and field-induced spin texture. While the plausibility of alter-magnetism in $CoTeMOO_6$ is conclusively ruled out by our single crystal neutron diffraction, more experimental research is required to explore the possible existence of field-induced exotic spin texture. Moreover, it is uncertain how the existence of strong in-plane anisotropy and twin domains further influences the underlying spin textures. Our experimental investigation and theoretical calculations indicate a significant interplay among magnetic, structural, electronic, and optical properties in $CoTeMoO_6$. In particular, the potential tunability of structural chirality with external parameters could be of considerable interest for both fundamental research and technical applications. Overall, these findings suggest that $CoTeMoO_6$, as well as other members of $ATeMoO_6$, could serve as interesting platforms for investigating complex physics with intertwined degrees of freedom.

**Acknowledgements**

Y.L. is grateful to F. Womack and S. Karna for the assistance with measurements. J.H.C. thanks A. Roy, for DSC measurements at the Center for Advanced Microstructure and Devices, LSU. We acknowledge D.M. Cao, Y. Mu at the shared Instrumentation Facility (SIF), LSU, for chemical and structural analysis. Work at Argonne National Laboratory was supported by the U.S. Department of Energy, Office of Science, Office of Basic Energy Sciences, Materials Sciences and Engineering Division. Crystal growth, magnetic and optical measurements were supported by the U.S. Department of Energy under EPSCoR Grant No. DESC0012432 with additional support from the Louisiana Board of Regents. The Neutron diffraction research used resources at the Spallation Neutron Source, a DOE Office of Science User Facility operated by Oak Ridge National Laboratory.

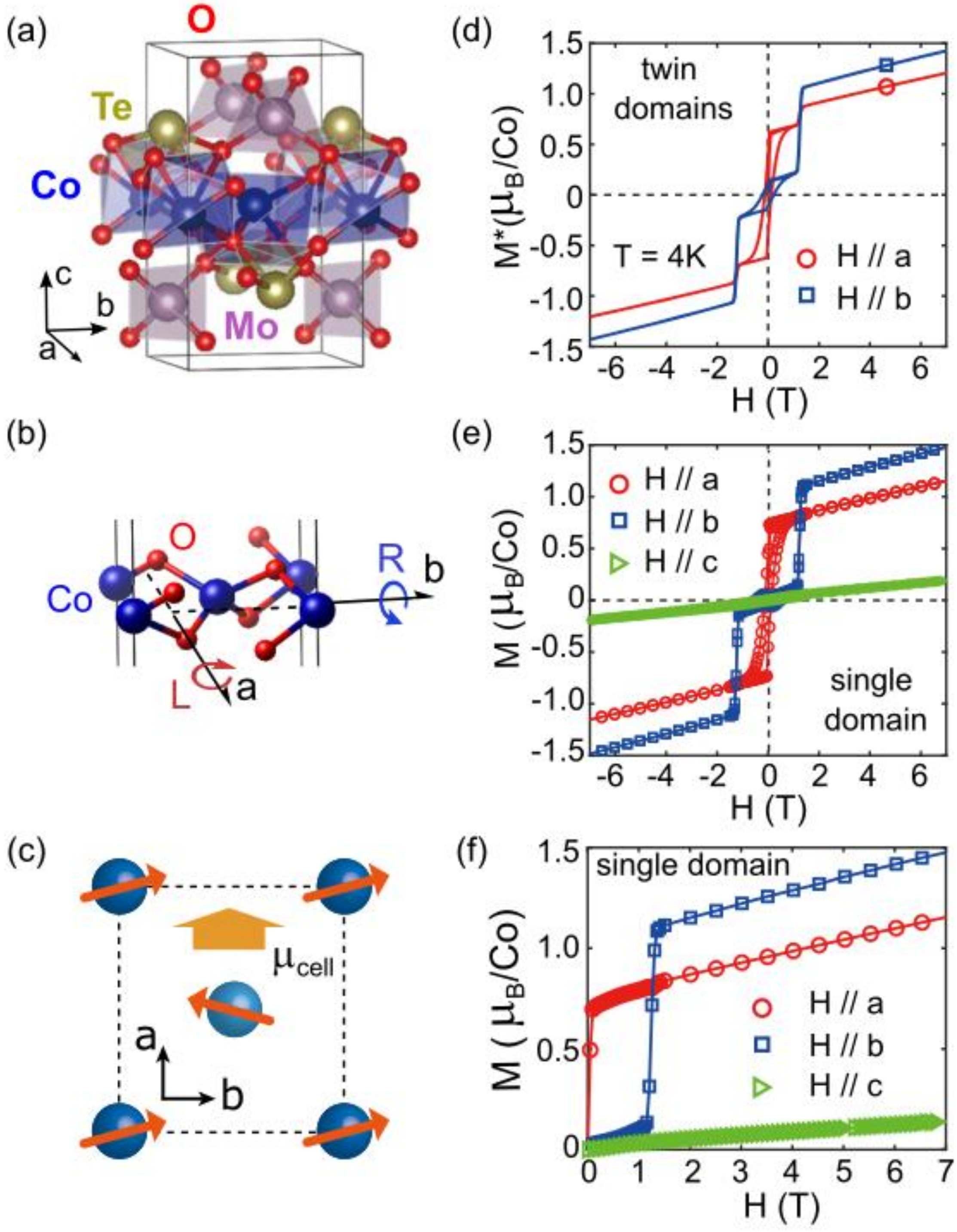

**Figure 1**. (a) The crystal structure of $CoTeMoO_6$. Co, Te, Mo, and O are represented by different colors, and Co ions are located in distorted octahedra that are connected by sharing one corner oxygen. (b) Illustration of the 2-fold helix of Co connected by Co-O bonds. (c) The arrangement of magnetic moments within the ab plane. (d) The measured magnetization, $M^*$, of a twinned $CoTeMoO_6$ with twinning domains at T = 4K. The magnetic field is applied along the nominal a (red) and b (blue) axis. We note here that the determination of the nominal a/b axis is based on the measurements of angle-

dependence and uniaxial pressure effect as discussed in the latter sections. (e) The extracted hysteresis curves for single domain, $M_x$, with $x = a, b, c$, as described in the text. (f) The extracted magnetization curves after zero-field-cooling (ZFC) for single domain, $M_x$, in which $x = a, b, c$.

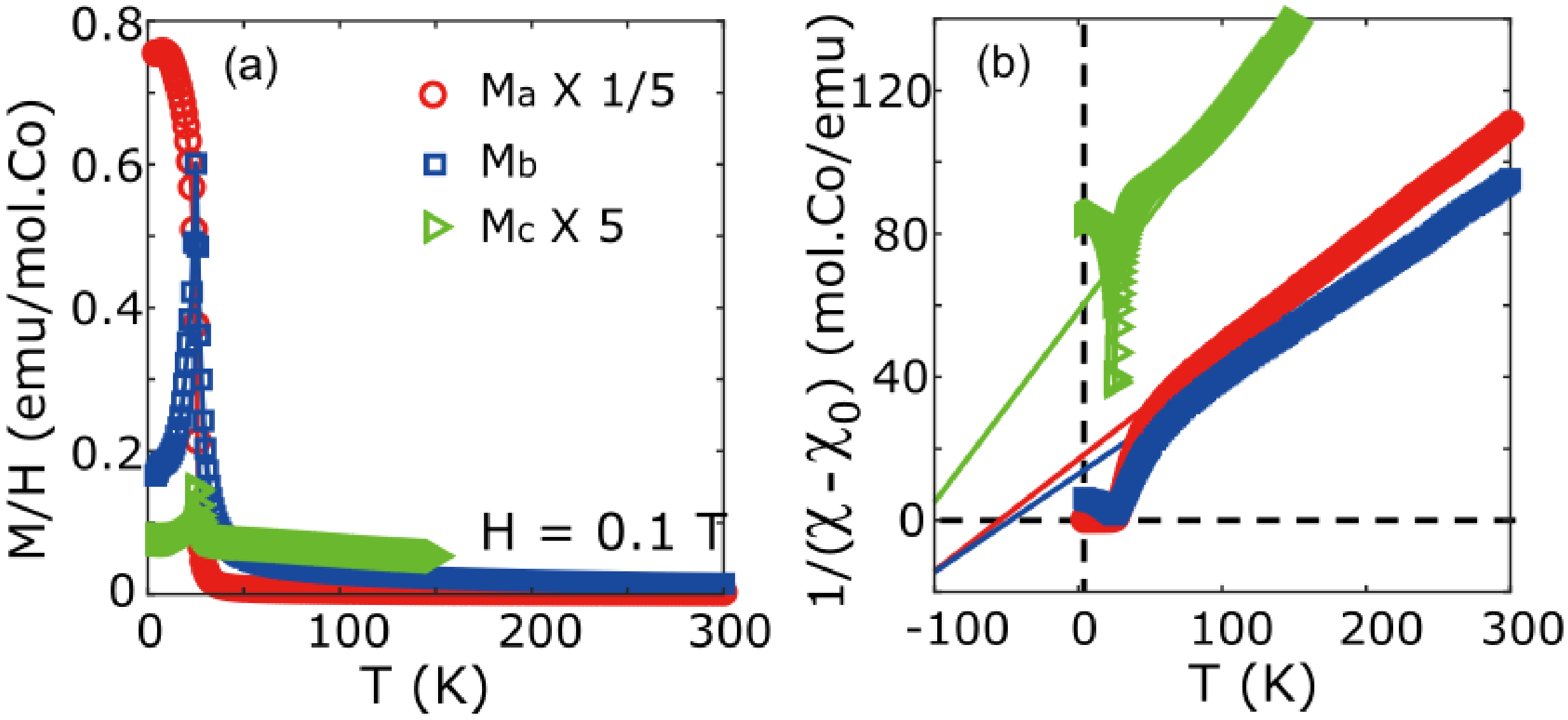

**Figure 2**. (a) Temperature dependence of zero-field-cooled magnetic susceptibility, $M/H$, of $CoTeMoO_6$ under a magnetic field of 0.1 T, after taking the twinning effect into consideration. (b) The inverse susceptibility curves along the a, b and c crystalline axes with solid curves representing Curie-Weiss fits of the corresponding data in panel (a).

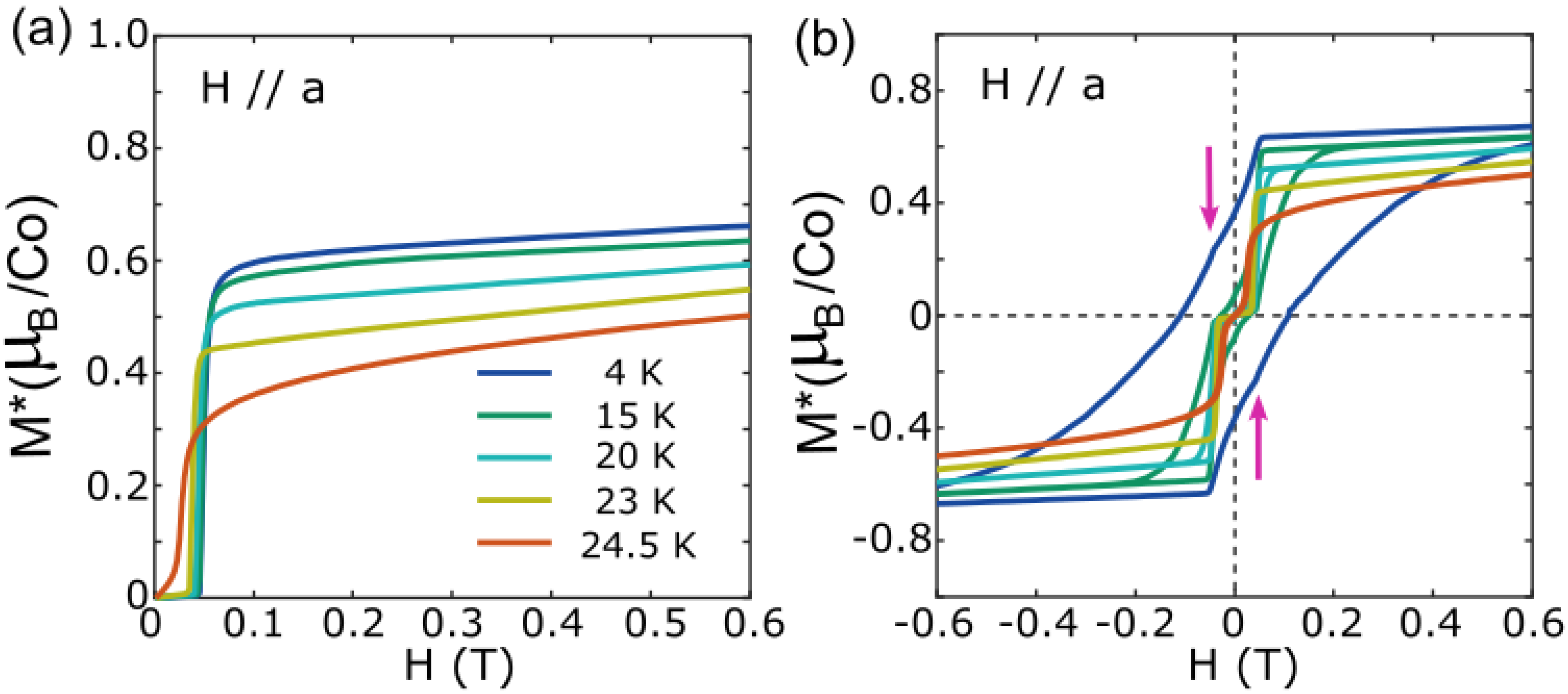

**Figure 3**. (a) Measured magnetization curves after ZFC vs magnetic field within the low-field region at varying temperatures. The magnetic field is applied along the crystalline ***a***-axis. A field-induced magnetic transition appears at approximately 0.05 T. (b) The magnetic hysteresis measured at the corresponding temperatures.

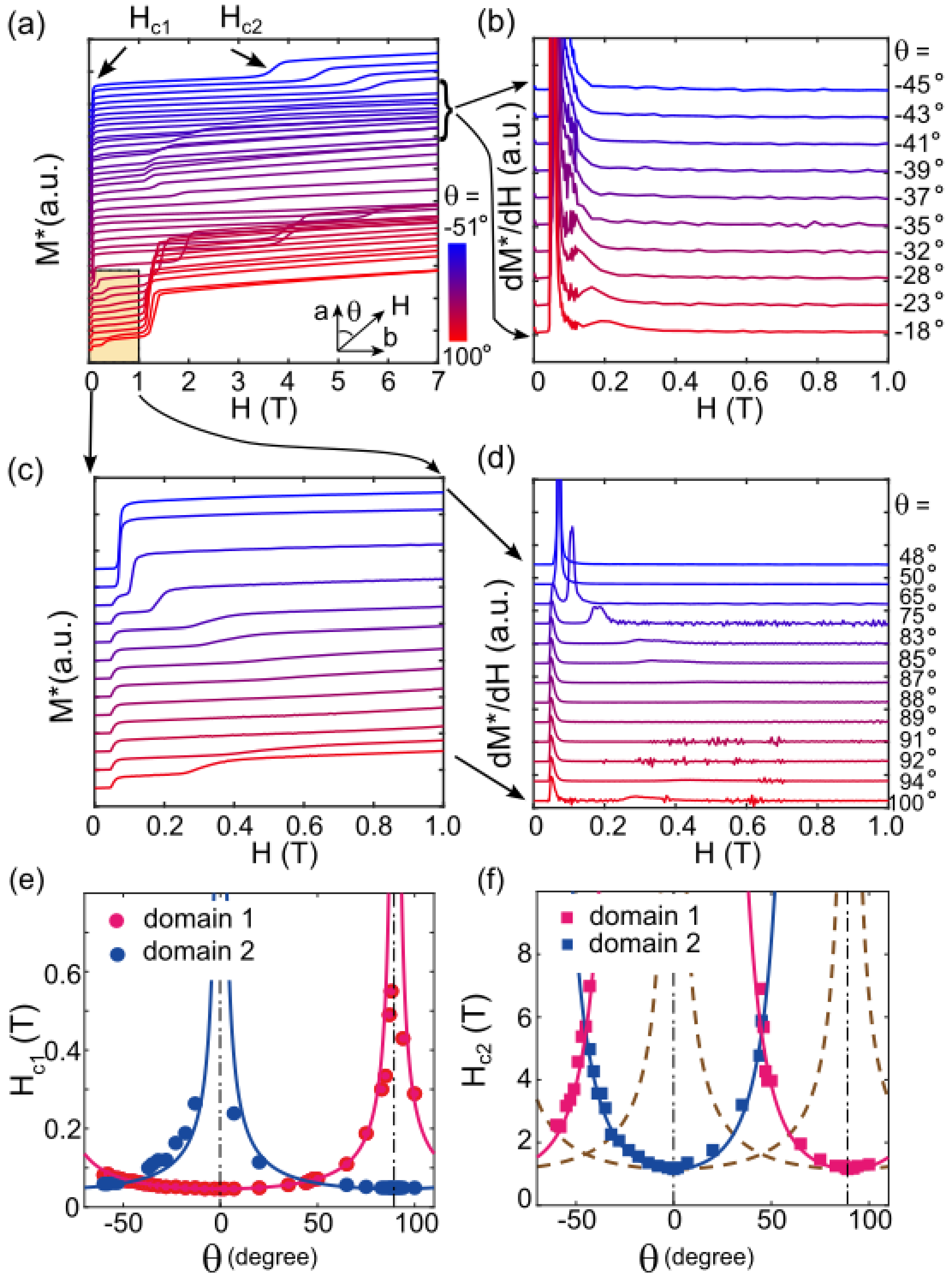

**Figure 4**. The angle dependence of magnetic behavior for $CoTeMoO_6$ measured at 2K. (a) The measured magnetization, $M^*$, at different orientations with $\theta$= -51°, -49°, -47°, -43°, -41°, -39°, -37°, -35°, -32°, -28°, -23°, -18°, -13°, -8°, -3°, 0°, 2°, 7°, 20°, 35°, 44°, 45°, 46°, 47°, 48°, 50°, 65°, 75°, 83°, 85°,

87°, 88°, 89°, 91°, 92°, 94°, and 100°, presented from the top to the bottom, where $\theta$ is the angle between the field and the nominal a-axis. (b) The first derivative of the measured magnetic susceptibility, $dM^*/dH$, with in the field range from 0 to 1 T, at selected orientations as indicated in (a). Beside the primary peaks, secondary peaks are clearly seen and exhibit angle dependence. (c) and (d) The measured magnetization, $M^*$, and the corresponding first derivative, $dM^*/dH$, at selected orientations, as enclosed in the black box in panel (a). The feature of $H_{c1}$ from a second domain is clearly seen. (e) The extracted angular dependence of $H_{c1}$. Blue and red colors represent the different twinning domains. The solid lines indicate the fit of the function $H_{c1} = H_{c1,0} \times \cos^{-1}(\theta - \theta_0)$. (f) The extracted angular dependence of $H_{c2}$. The solid curves fit a function of $H_{c2} = H_{c2,0} \times \cos^{-4.5}(\theta - \theta_0)$, with an exponent of -4.5 instead of -1. For comparison, the function of $H_{c2,0} \times \cos^{-1}(\theta - \theta_0)$ is plotted as dashed curves.

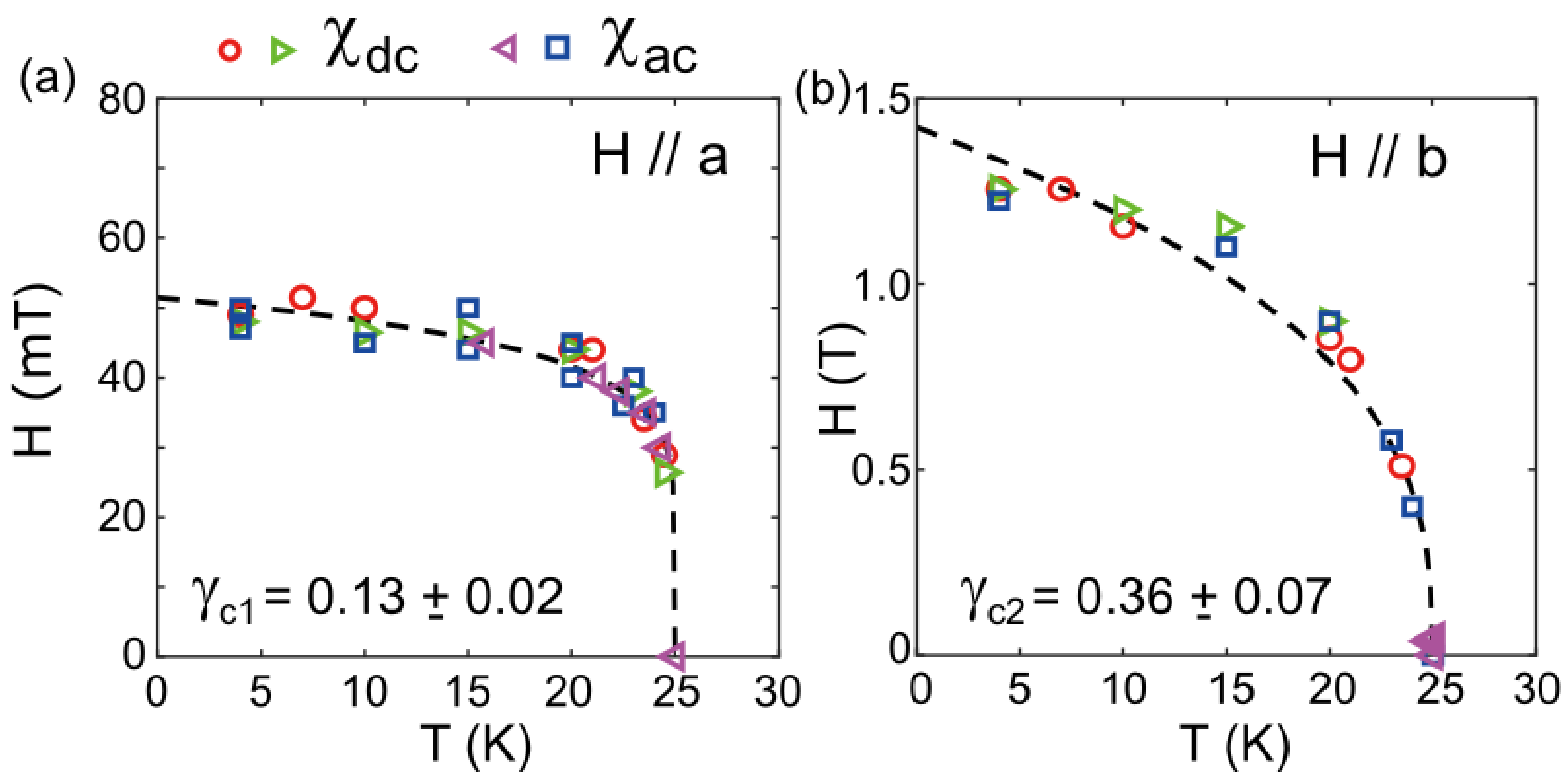

Figure 5. The H-T phase diagram of $CoTeMoO_6$ under applied fields along the a-axis in the panel (a), and the b-axis in panel (b), separately. The different markers represent the corresponding extracted values from different measurements. The dashed curves are fit of a function of $H_c = H_0 \times (1 - T/T_c)^{\gamma}$ in which $T_c$ is fixed at 25 K in the fitting process for both data sets. It is noteworthy that the exponents for $H_{c1}$ and $H_{c2}$ are different.

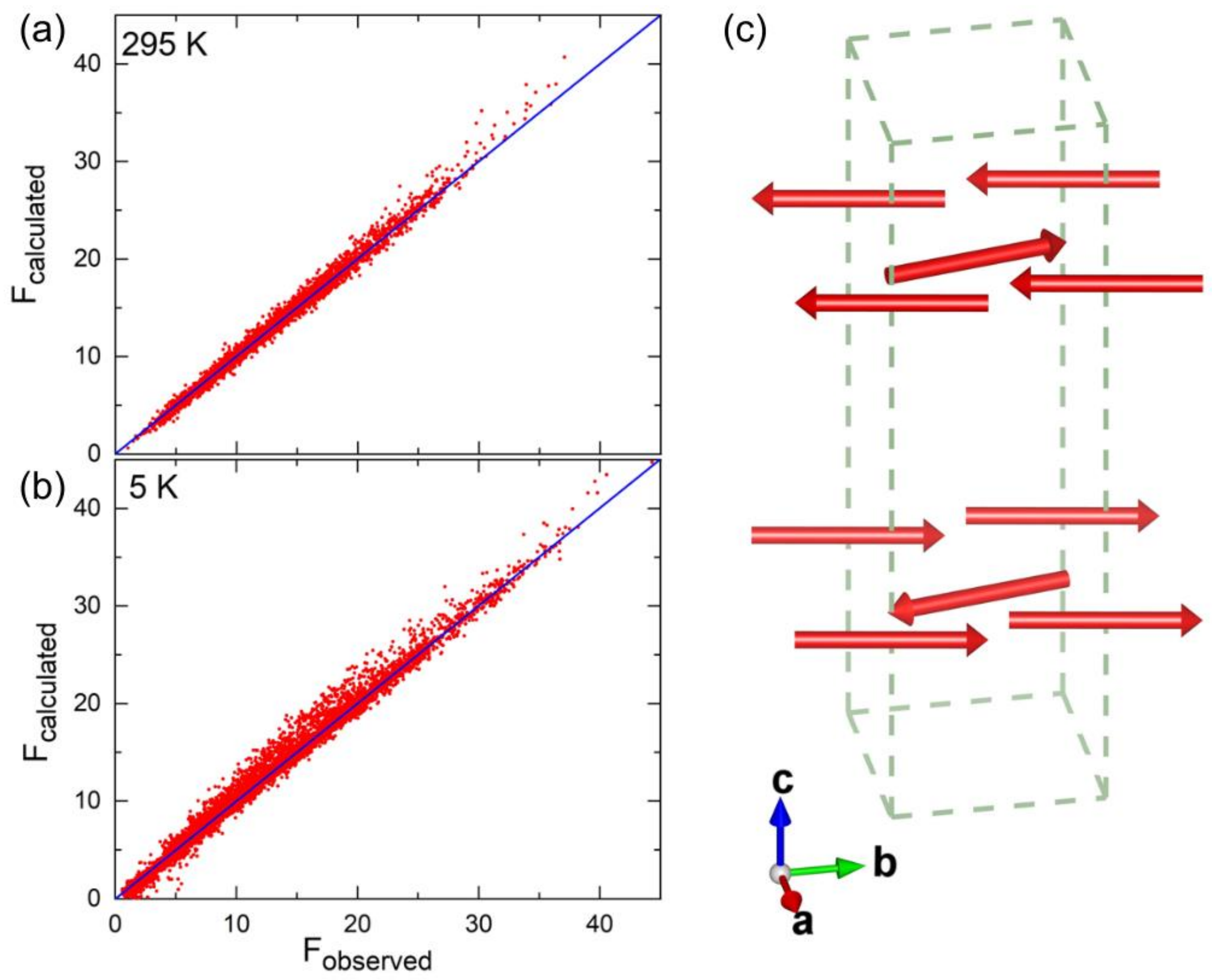

Figure 6. (a), (b) The results from the refinement of the structure at 295 K and 5 K, respectively. The agreement between the calculated and observed values is reasonably well. (c) The refined magnetic structure from single crystal neutron diffraction at 5 K.

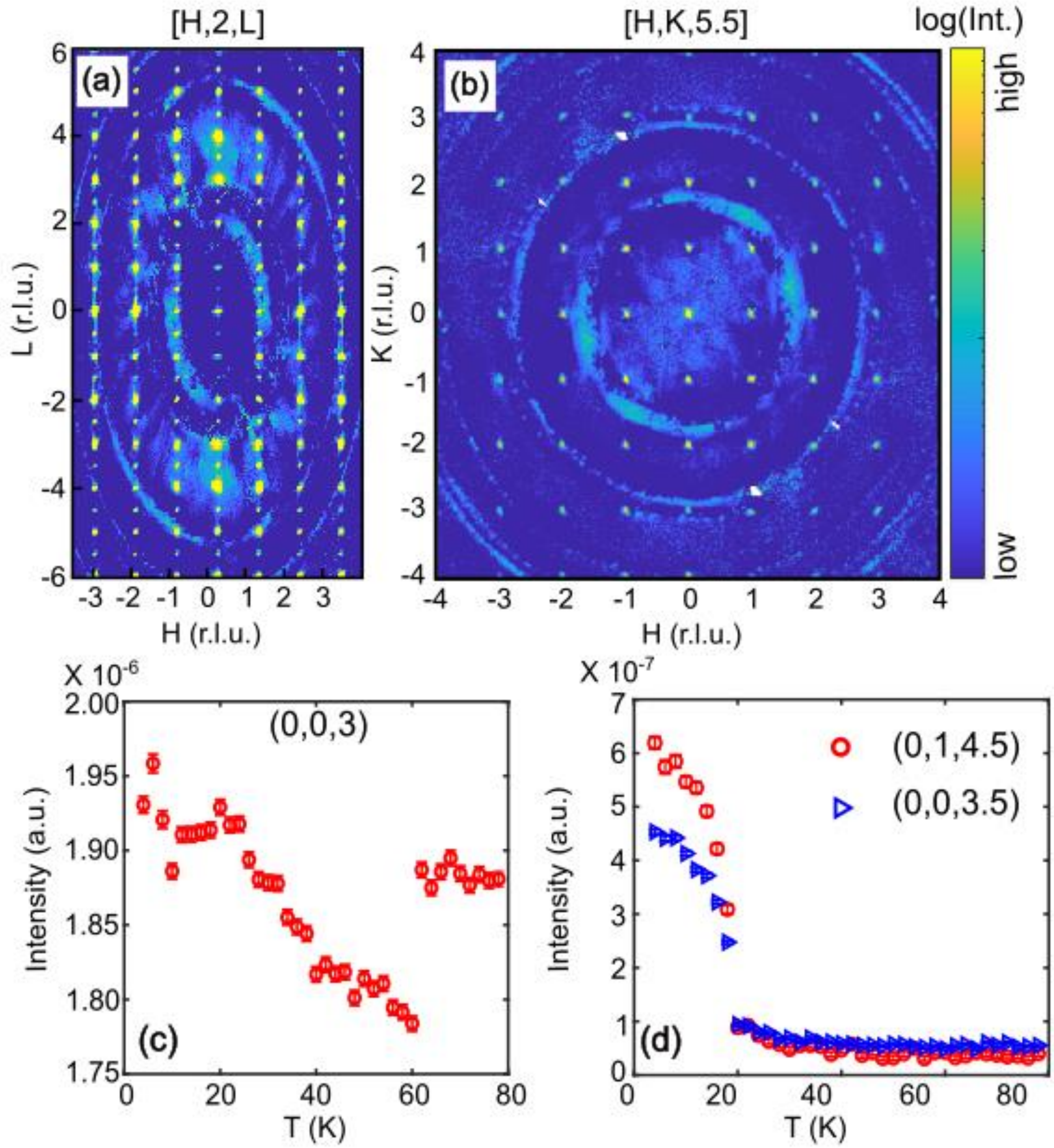

Figure 7. (a) A Neutron diffraction image of $CoTeMoO_6$ at 5 K is illustrated within the $[H, 2, L]$ plane where the magnetic peaks can be observed at half-integer L values. (b) A image of neutron diffraction intensities within the $HK$ plane with $L = 5.5$ at 5 K. (c) The temperature dependence of neutron diffraction intensity at (0,0,3). (d) The temperature dependence of neutron diffraction intensity at $Q_{AFM}$ = (0,0,3.5) and (0,1,4.5).

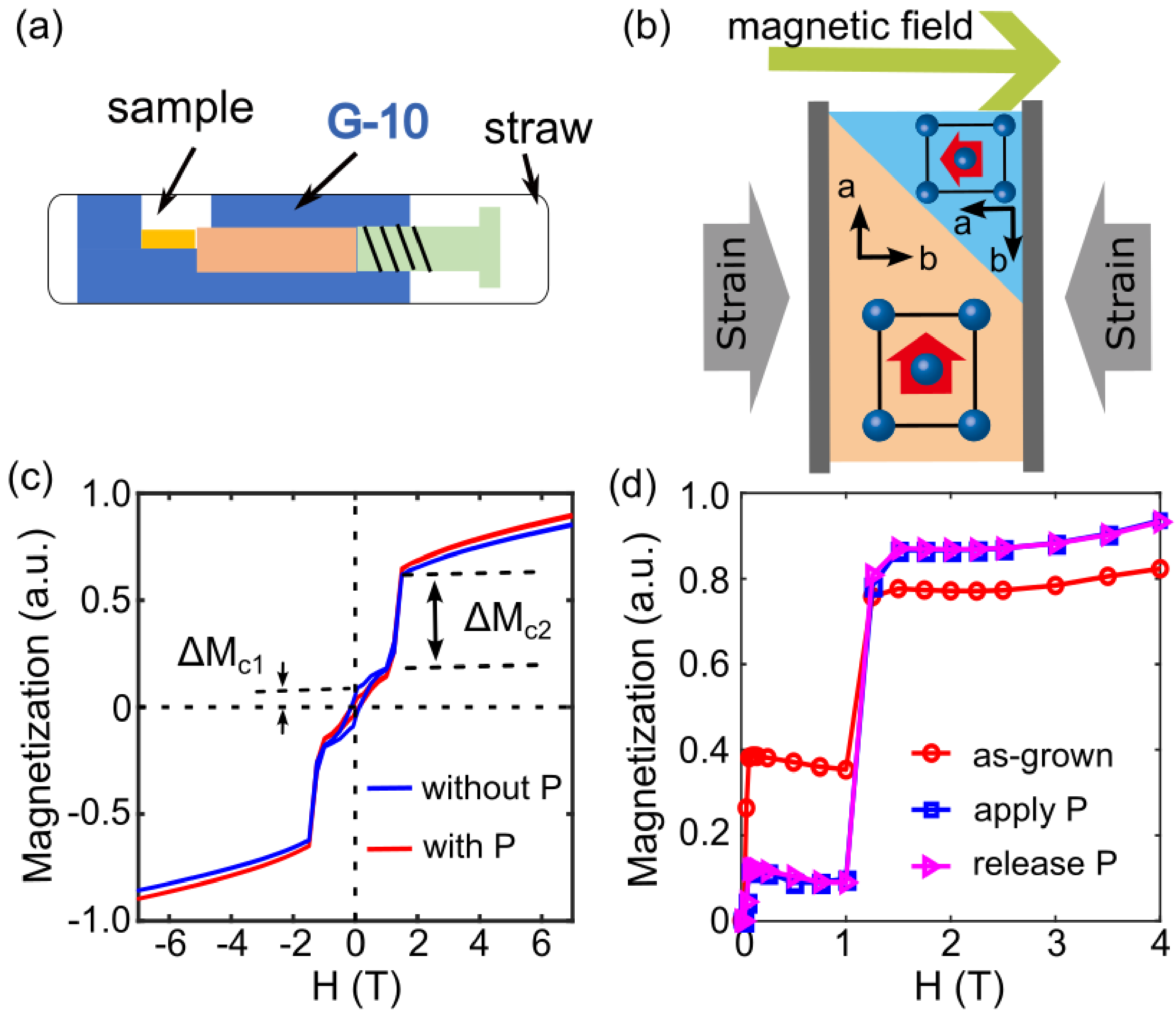

Figure 8. (a) Schematics of a homemade pressing device used for magnetization measurements in MPMS. (b) An illustration of the detwinning process in the sample under uniaxial pressure. (c) The measured magnetization curves with (red) and without (blue) uniaxial pressure at 4 K. The background signal from an empty device was measured and subtracted. (d) The ZFC magnetization of the sample in pristine condition (red), with a uniaxial pressure applied (blue), and then after releasing the pressure (magenta). The sample was taken out from MPMS at room temperature for the application and release of pressure.

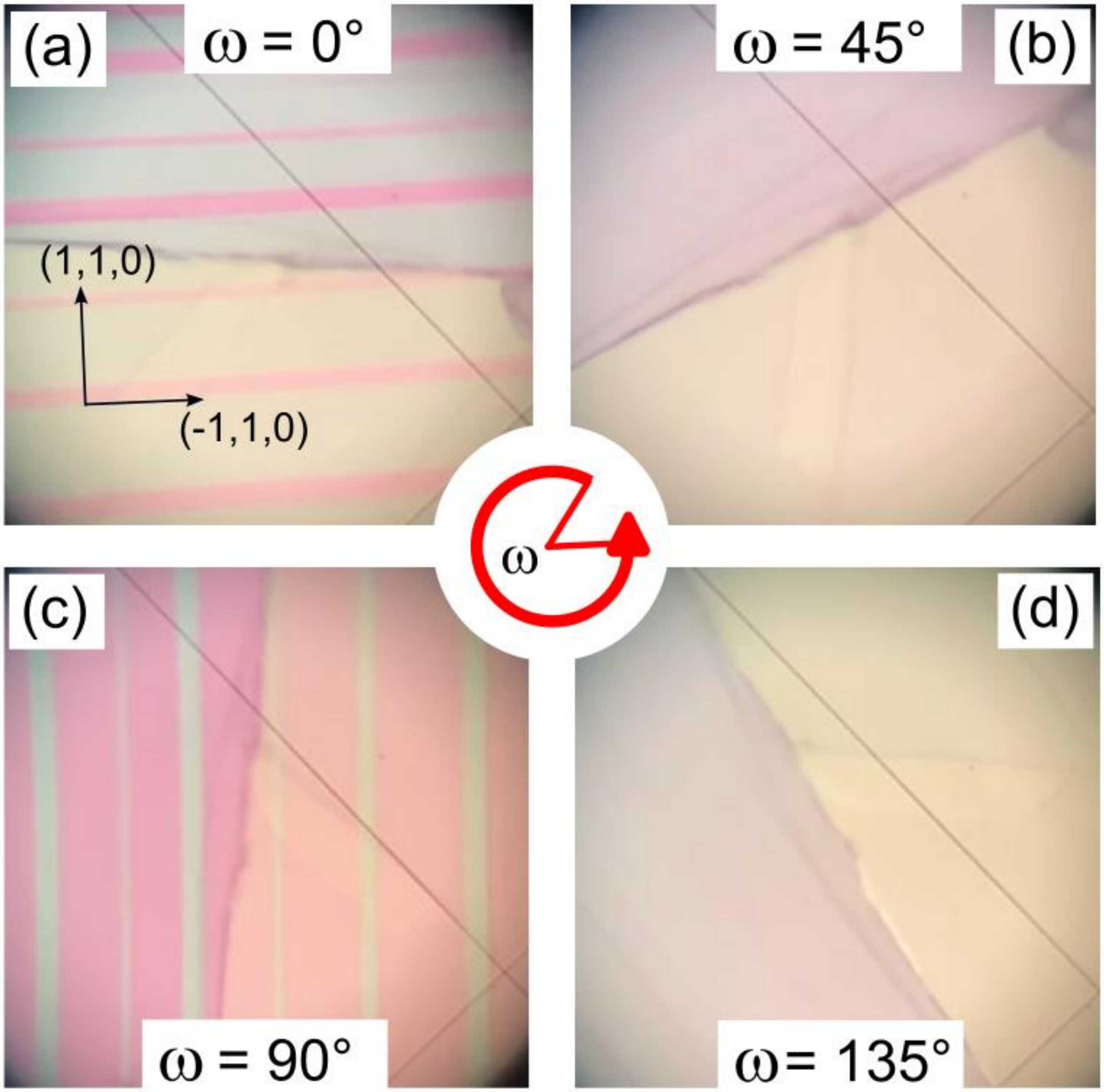

Figure 9. Images of a crystal of $CoTeMoO_6$ under observation using an optical microscope. A light polarizer was placed between the sample and the light source, revealing an anisotropic response to polarized light when the sample was rotated with respect to the azimuthal direction $\omega$. Stripe-like domains are clearly visible in (a) and (c), while the contrast completely disappears in (b) and (d). It is noteworthy that the contrast of stripe domains in (a) and (c) is reversed.

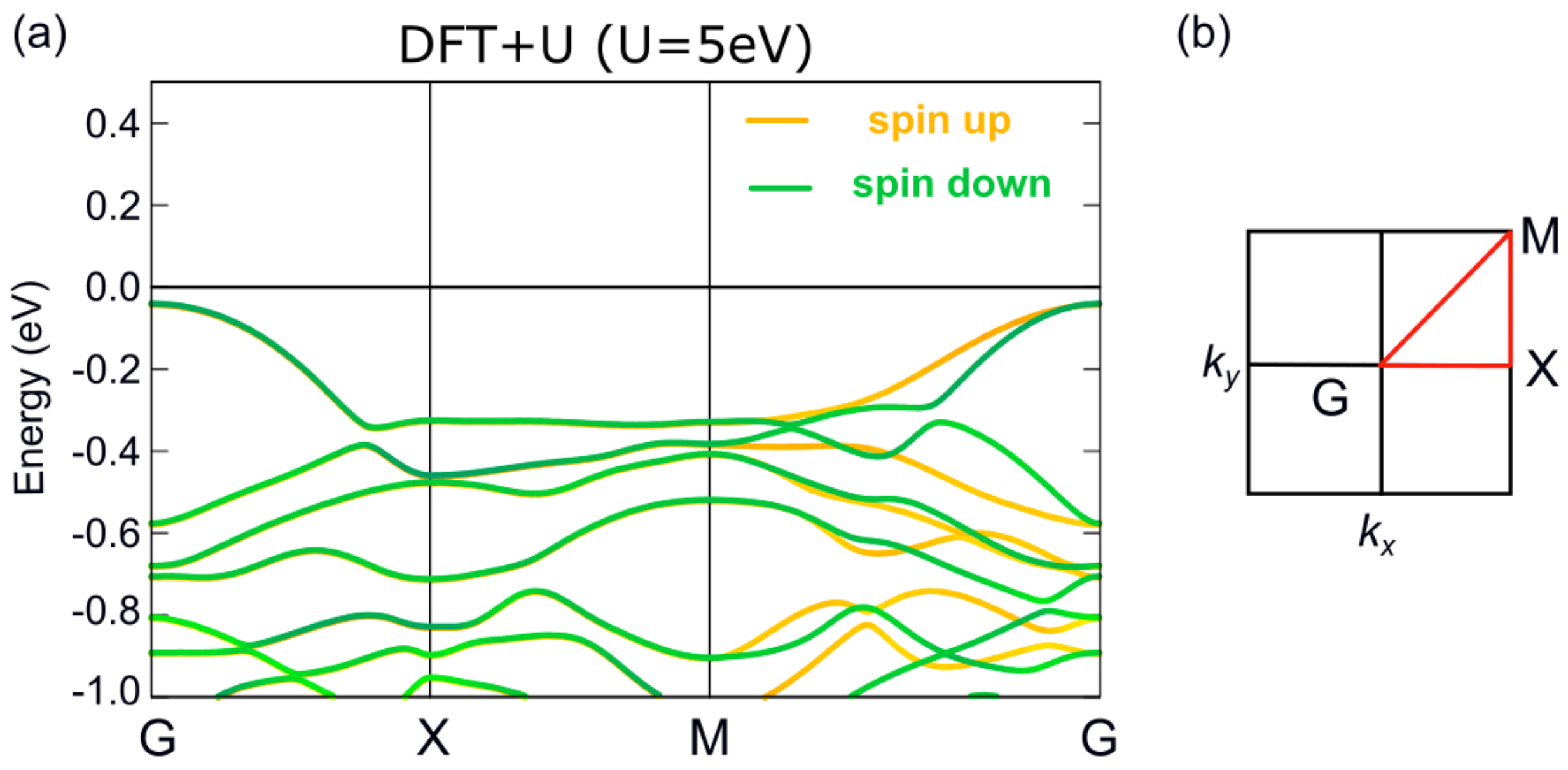

Figure 10. (a) The *Ab initio* spin splitting of the band structure in $CoTeMoO_6$, which was obtained with U = 5 eV and an AFM spin arrangement between the two Co ions within the structural unit cell. The band splitting between the up and down spins is clearly seen along the G-M direction. (b) An illustration of the k-in the 2D Brillouin Zone (BZ).

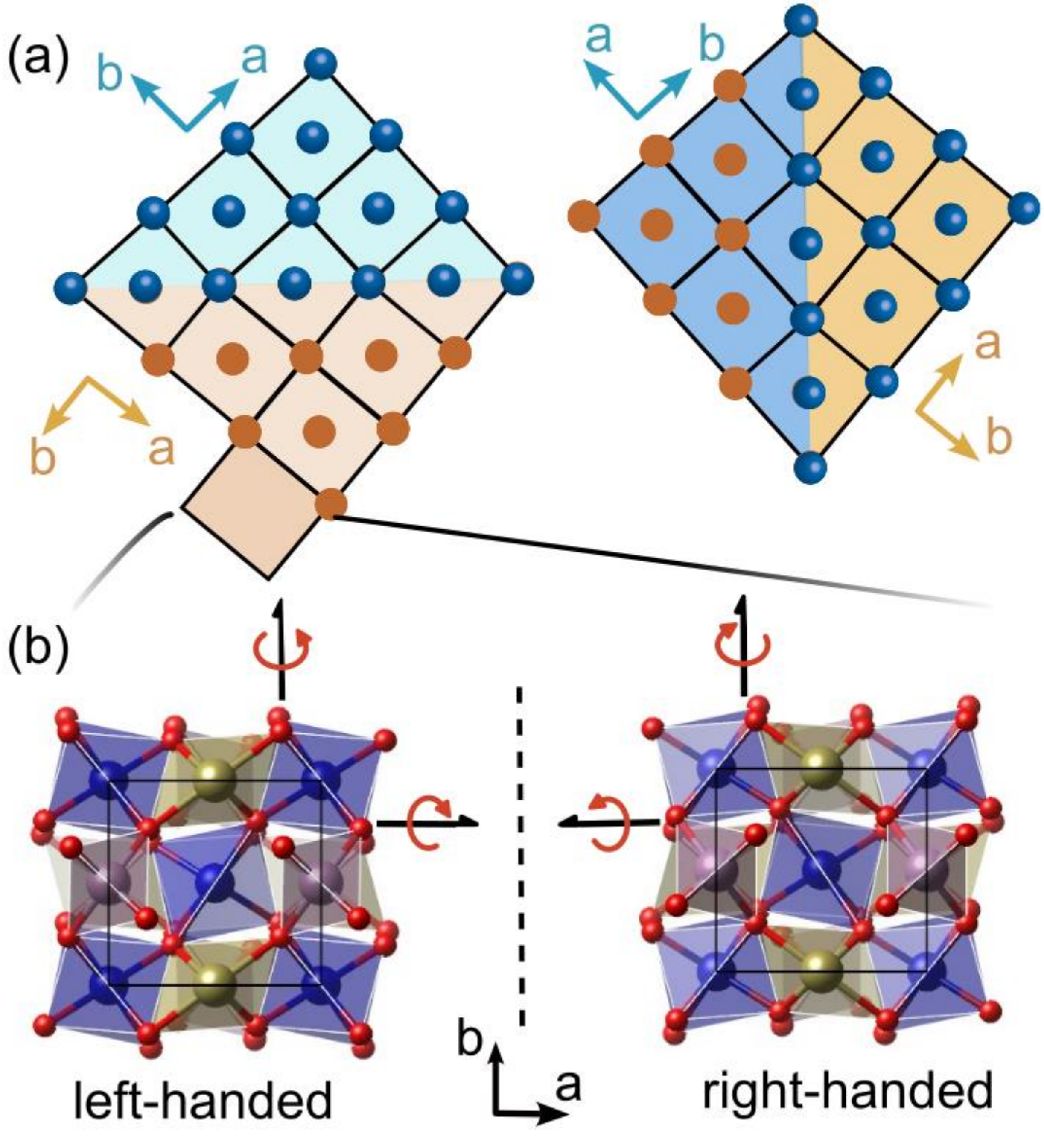

Figure 11. Illustration of the twin domains in $CoTeMoO_6$. (a) Coexistence of orthorhombic domains with four different orientations. (b) Enantiomers of $CoTeMoO_6$ with respect to one orthorhombic unit cell. The transformation from one enantiomer to the other involves the rotation and distortion of $CoO_6$ octahedra, $MoO_4$ tetrahedra and $TeO_4$ polyhedra.

**Table 1.** Single crystal refinement for $CoMoTeO_6$ at 293 (2) K from X-ray diffraction.

| Refined Formula | $CoMoTeO_6$ |
|---|---|
| F.W. (g/mol) | 378.47 |
| Space group; Z | $P\ 2_12_12$; 2 |
| $a$ (Å) | 5.2457 (1) |
| $b$ (Å) | 5.0572 (1) |
| $c$ (Å) | 8.8403 (2) |
| V ($Å^3$) | 234.520 (8) |
| θ range (deg) | 2.304-40.260 |
| No. reflections; $R_{int}$ | 7017; 0.0463 |
| No. independent reflections | 1487 |
| No. parameters | 44 |
| $R_1$: $\omega R_2$ ($I$>2δ($I$)) | 0.0264; 0.0525 |
| Goodness of fit | 1.056 |
| Diffraction peak and hole ($e^-$/ $Å^3$) | 1.479; -1.216 |
| Absolute structure parameter | 0.05 (5) |

**Table 2.** Atomic coordinates and equivalent isotropic displacement parameters of $CoMoTeO_6$ from X-ray difraction. ($U_{eq}$ is defined as one-third of the trace of the orthogonalized $U_{ij}$ tensor ($Å^2$))

| **Atom** | **Wyck.** | **Occ.** | ***x*** | ***y*** | ***z*** | $U_{eq}$ |
|---|---|---|---|---|---|---|
| Te1 | 2*a* | 1 | 0 | ½ | 0.25405 (3) | 0.0067 (1) |
| Mo2 | 2*a* | 1 | 0 | ½ | 0.80691 (5) | 0.0085 (1) |
| Co3 | 2*b* | 1 | 0 | 0 | 0.52577 (8) | 0.0090 (1) |
| O4 | 4*c* | 1 | 0.2203 (6) | 0.1778 (6) | 0.1083 (4) | 0.0106 (5) |
| O5 | 4*c* | 1 | 0.2408 (6) | 0.6826 (7) | 0.2039 (4) | 0.0123 (6) |
| O6 | 4*c* | 1 | 0.3469 (7) | 0.2891 (7) | 0.4250 (4) | 0.0142 (6) |

**Table 3.** Anisotropic thermal displacements from $CoMoTeO_6$ from X-ray diffraction.

| Atom | U11 | U22 | U33 | U12 | U13 | U23 |
|---|---|---|---|---|---|---|
| Te1 | 0.0060 (1) | 0.0059 (1) | 0.0082 (1) | 0.0004 (3) | 0 | 0 |
| Mo2 | 0.0087 (2) | 0.0068 (2) | 0.0100 (2) | -0.0032 (4) | 0 | 0 |
| Co3 | 0.0080 (2) | 0.0069 (2) | 0.0120 (3) | -0.0009 (6) | 0 | 0 |
| O4 | 0.008 (1) | 0.010 (1) | 0.014 (1) | 0.002 (1) | -0.003 (1) | -0.005 (1) |
| O5 | 0.009 (1) | 0.012 (1) | 0.017 (2) | 0.005 (1) | 0.001 (1) | 0.001 (1) |
| O6 | 0.015 (1) | 0.014 (1) | 0.014 (2) | -0.003 (1) | -0.002 (1) | 0.001 (1) |

**Table 4.** Data collection parameters for neutron diffraction.

| | | 295 K | 5 K |
|---|---|---|---|
| **Number of Reflections** | Total<br>Nuclear<br>Magnetic | 4612<br>4612<br>--- | 7917<br>5934 (nuclear)<br>1983 (magnetic) |
| ***h, k, l* limits** | | $-10 \le h \le 7$<br>$-9 \le k \le 9$<br>$-17 \le l \le 8$ | $-9 \le h \le 9$<br>$-9 \le k \le 9$<br>$-16 \le l \le 16$ |
| **Minimum Intensity (I)** | | > 3 sig(I) | > 3 sig(I) |
| **Min theta, deg.** | | 5.08° | 1.91° |
| **Max theta, deg.** | | 53.2° | 32.41° |

**Table 5.** Structural Parameters for $CoMoTeO_6$ from neutron diffraction.

| **T (K)** | | **295 K** | **5 K** |
|---|---|---|---|
| **Nuclear Space Group** | | $P2_12_12$ | $P2_12_12$ |
| **Magnetic Space Group** | | | $P_C2_12_12_1$ |
| ***a*** **(Å)** | | 5.2464 | 5.2442 |
| ***b*** **(Å)** | | 5.0579 | 5.0592 |
| ***c*** **(Å)** | | 8.8431 | 17.6576 |
| $\alpha = \beta = \gamma$ | | 90° | 90° |
| ***Volume*** **(Å³)** | | 234.7 | 468.5 |
| **Co @ (0, 0, z)** | z | 0.52602(8) | 0.26330(3) |
| | $U_{iso}$ (Å$^2$) | 0.55(1) | 0.040(4) |
| | Mx | | 0.666(7) |
| | My | | 3.379(7) |
| | Mz | | 0 |
| | M | | 3.444(7) |
| **Te @ (0, ½ z)** | z | 0.25396(3) | 0.12655(2) |
| | $U_{iso}$ (Å$^2$) | 0.394(6) | 0.040(4) |
| **Mo @ (0, ½, z)** | z | 0.80709(3) | 0.40388(2) |
| | $U_{iso}$ (Å$^2$) | 0.505(6) | 0.040(4) |
| **O1 @ (x, y, z)** | x | 0.28974(8) | 0.28888(6) |
| | y | 0.84606(8) | 0.84635(7) |
| | z | 0.07484(3) | 0.03670(2) |
| | $U_{iso}$ (Å$^2$) | 1.051(7) | 0.292(6) |
| **O2 @ (x, y, z)** | x | 0.18355(7) | 0.18316(7) |
| | y | 0.75851(7) | 0.75963(7) |
| | z | 0.70389(3) | 0.35201(2) |
| | $U_{iso}$ (Å$^2$) | 0.761(5) | 0.239(6) |
| **O3 @ (x, y, z)** | x | 0.17803(7) | 0.17857(7) |
| | y | 0.71977(6) | 0.72097(7) |
| | z | 0.39145(3) | 0.19550(2) |
| | $U_{iso}$ (Å$^2$) | 0.683(5) | 0.220(6) |
| $R_{obs\ nuclear}$ **(%)** | | 3.19 | 4.48 |
| $wR_{obs\ nuclear}$ **(%)** | | 8.43 | 4.40 |
| $R_{obs\ magnetic}$ **(%)** | | | 11.09 |
| $wR_{obs\ magnetic}$ **(%)** | | | 7.67 |
| **Goodness of Fit (GOF)** | | 1.04 | 3.18 |
| | | | |

**Table 6.** Bond-lengths and Bond Angles at 295K and 5K from neutron diffraction.

| Bond-Lengths and Bond Angles | | 295 K | 5 K |
|---|---|---|---|
| Co-O3 (Å) | x 2 | 2.0730(5) | 2.0743(4) |
| Co-O3 (Å) | x 2 | 2.1498(4) | 2.1495(4) |
| Co-O2 (Å) | x 2 | 2.2121(6) | 2.2034(5) |
| O3-Co-O3 (°) | x 2 | 101.21(1) | 101.26(1) |
| O3-Co-O3 (°) | x 2 | 101.27(1) | 101.27(1) |
| O3-Co-O3 (°) | x 1 | 109.96(3) | 109.50(3) |
| O3-Co-O3 (°) | x 1 | 140.31(4) | 140.44(3) |
| O2-Co-O3 (°) | x 2 | 169.06(3) | 169.32(2) |
| | | | |
| Te-O3 (Å) | x 2 | 1.8938(4) | 1.8997(4) |
| Te-O2 (Å) | x 2 | 2.0946(4) | 2.0935(4) |
| O3-Te-O3 (°) | x 1 | 100.10(2) | 100.29(2) |
| O2-Te-O2 (°) | x 1 | 159.50(2) | 159.17(2) |
| | | | |
| Mo-O1 (Å) | x 2 | 1.7069(4) | 1.7121(4) |
| Mo-O2 (Å) | x 2 | 1.8627(4) | 1.8673(4) |
| O1-Mo-O1 (°) | x 1 | 104.57(2) | 104.40(2) |
| O1-Mo-O2 (°) | x 2 | 108.28(2) | 108.29(2) |
| O1-Mo-O2 (°) | x 2 | 106.61(2) | 106.70(2) |
| O2-Mo-O2 (°) | x 1 | 121.33(2) | 121.26(2) |